\documentclass[preprint]{aastex}

\def\ltsima{$\; \buildrel < \over \sim \;$}

\def\lsim{\lower.5ex\hbox{\ltsima}}
\def\gtsima{$\; \buildrel > \over \sim \;$}
\def\gsim{\lower.5ex\hbox{\gtsima}}

\begin{document}
\title{Cosmological Microlensing Statistics: Variability rates for Quasars and GRB Afterglows, and implications for macrolensing magnification bias and flux ratios}

\author{J. S. B. Wyithe\altaffilmark{1,3}, E. L. Turner\altaffilmark{2}}

\altaffiltext{1}{Harvard College Observatory, 60 Garden St, Cambridge, MA 02138, USA}

\altaffiltext{2}{Princeton University Observatory, Peyton Hall, Princeton, NJ 08544, USA}

\altaffiltext{3}{Hubble fellow}

\begin{abstract}

The fraction of quasar's and gamma-ray burst (GRB) afterglows that vary due to microlensing by the stellar populations of intervening elliptical/S0 galaxies is computed by combining the joint distribution of effective microlensing convergence ($\kappa$) and shear ($\gamma$) with microlensing magnification patterns. Microlensing is common in multiply imaged sources. We find that 1 in 3 multiply imaged quasars should vary by more than 0.5 magnitudes per decade due to microlensing, while 10\% of macrolensed GRB afterglows should show a departure of more than 0.5 magnitudes from their intrinsic light-curve during the first 30 days. However microlensing by stars is rare in general, with only 1 source in $\sim500$ varying by more than 0.5 magnitudes during the same periods. We find that most microlensing by stars will be observed in a regime where $\gamma>0.1$. Thus point-mass lenses do not provide an adequate description for most microlensing events. If dark matter halos contain a large fraction of mass in compact objects, the fraction of microlensed (by 0.5 magnitudes) images rises significantly to $\sim1$ in 10 for quasars and $\sim$1 in 5 for GRB afterglows. Comparison of variability between macrolensed and normal quasar images, and a moderate number of well sampled GRB afterglow light-curves should therefore discover or refute the existence of stellar mass compact objects in galaxy halos. While microlensing results in departures of the distribution of magnifications from that of a smooth model, the effect on the macrolensing magnification bias for the discovery of lenses in quasar surveys is small. On the other hand, microlensing significantly broadens the distribution of macrolensed image flux ratios.

\end{abstract}


\section{Introduction}

Cosmological microlensing of background quasars by stars in foreground galaxies was first discussed in the work of Chang \& Refsdal~(1979), Gott~(1981) and Young~(1981). Since then considerable theoretical and observational progress has been made towards using microlensing phenomenon to obtain otherwise inaccessible information on the cosmological distribution of stellar and sub-stellar mass compact objects, and on the smallest scales of the central engines of quasars. In contrast, there has been relatively little work (Bartelmann \& Schneider~1990; Koopmans \& Wambsganss~2000; Wyithe \& Turner~2002, hereafter Paper I) extending calculations of the a-priori probability for cosmological microlensing beyond the seminal work of Press \& Gunn~(1973).

Microlensing was first identified in Q2237+0305 (Irwin et al.~1989; Corrigan et al.~1990). Early models presented by Wambsganss, Paczynski \& Katz~(1989) and Wambsganss, Paczynski \& Schneider~(1990) explained the observed flux variation using a model containing stellar mass objects. Following this success, microlensing of cosmological sources (mostly quasars) has been used to discuss the nature of the intervening compact object populations (e.g. Press \& Gunn~1973; Blaes \& Webster~1992; Schneider 1993; Dalcanton et al.~1994; Lewis \& Irwin 1995; Perna \& Loeb~1997; Schmidt \& Wambsganss~1998; Marani et al.~1999; Wyithe, Webster, Turner \& Mortlock~2000; Wambsganss et al.~2000; Refsdal, Stabell, Pelt \& Schild~2000; Schild~1999; Mao \& Loeb~2001; Lewis \& Ibata~2001; Lewis, Ibata \& Wyithe~2000). Microlensing has also been used to constrain different physical emission regions of quasars. Variability of the optical continuum in Q2237+0305 (Corrigan et al.~1989; Irwin et al.~1990; $\O$stensen et al.~1996; Wosniak et a.~2001a,b) has been used to show that the emission region is $<10^{15}$cm (Wambsganss, Paczynski \& Schneider~1992; Rauch \& Blandford 1992; Jaroszynski, Paczynski \& Schneider~1992; Wyithe, Webster \& Turner~2000). Similarly, data for Q0957+561 (Pelt, Schild, Refsdal \& Stabell~1998) was used to find a source smaller than $5\times10^{15}$cm (Refsdal, Stabell, Pelt \& Schild~2000). On larger scales, Lewis et al.~(1998) observed microlensing induced variation between the optical continuum and broad-line regions, and Wyithe, Agol \& Fluke~(2002) used mid-IR to V-band flux ratios to demonstrate that the mid-IR emission comes from a region $>10^{17}$cm (based on the mid-IR data of Agol, Jones \& Blaes~1999). While these results are specific to Q2237+0305 (particularly) and Q0957+561 both of which have long monitoring histories, monitoring programs for the determination of time-delays (e.g. Kundic et al.~1997; Schechter et al.~1997; Burud et al.~2000) in multiply imaged quasars offer a promising avenue for more microlensing observations. For example $\O$stensen et al.~(1997) found evidence for low-level microlensing in monitoring of the Clover-leaf quasar.

More recently, additional novel microlensing observations have been published. These include evidence for microlensing of radio emission substructure in B1600+434 (Koopmans et al.~2000; Koopmans \& de Bruyn~2000), evidence for microlensing induced polarization variability in H1413+1143 (Chae et al.~2001), and evidence for microlensing of the Fe K$\alpha$ line in MG J0414+0534 (Chartas et al. 2001). Several other potential claims of AGN microlensing have also been published (e.g. Lewis, Robb \& Ibata~1999; Webb et al.~2000; Torres, Gustavo \& Eiroa~2002). 

In addition to the above lists of observations and interpretations, many studies have pointed out the potential value of detailed microlensing observations (particularly of caustic crossings). The most exciting are schemes to determine the emission spectrum of the central engine as a function of radius (e.g. Grieger, Kayser \& Refsdal~1988; Grieger, Kayser \& Schramm~1991; Agol \& Krolik~1999; Mineshige \& Yonehara~1999) from multiband observations of caustic crossings. In other examples, Schneider \& Wambsganss~(1990) showed that the structure and kinematics of Broadline regions could be constrained using observations of microlensed quasars, Belle \& Lewis~(2000) have demonstrated that microlensing should result in the variability of polarized light, and Lewis, Ibata \& Wyithe~(2000) and Lewis \& Ibata~(2001) have shown that cosmological compact objects could be detected through surface brightness variations in giant cluster arcs and galaxies. 

Quasars are not the only cosmological sources subject to microlensing variability. Another candidate is the GRB afterglow (Loeb \& Perna~1998). Indeed, Mao \& Loeb~(2001) find that all afterglows should be microlensed at a low level. Recently, a deviation from the generic double powerlaw decay was observed in monitoring of the afterglow light-curve of GRB000301C (Garnavich, Loeb \& Stanek~2000, based on data compiled by Sagar et al.~2000) and interpreted as due to microlensing. Detailed event inversions modeled after work on the interpretation of quasar caustic crossing events (e.g. Geiger et al.~1988) have since been performed (Gaudi, Granot \& Loeb 2001), and the lens hypothesis found to be plausible. Furthermore, Koopmans \& Wambsganss~(2000) have discussed the a-posteriori probability for microlensing of this source and found that while unlikely, the microlensing scenario cannot be ruled out. Their calculation went two steps beyond the formalism of Press \& Gunn~(1973). Firstly, they assumed microlenses to be clustered in galaxies which were modeled as SIS's of compact objects. Secondly, they computed microlensing statistics using magnification patterns rather than a point mass lens approximation.

Microlensing of high redshift type Ia supernovae has been discussed by Metcalf \& Silk~(1999) and Wang~(1999), and latter by M$\ddot{\mbox{o}}$rtsell, Goodbar \& Bergstr$\ddot{\mbox{o}}$m~(2001) and Minty, Heavens \& Hawkins~(2001) with emphasis on the utility of forthcoming survey samples. Type Ia supernovae are particularly interesting microlensing sources because of their status as standard candles. They are not expected to act as point sources for the typical duration of observations, and the resulting microlensing induced departures from the intrinsic light-curve will impact on the interpretation of the light-curve shape, and thus the inference of the intrinsic brightness. A study of microlensing and the light-curves of type Ia supernovae based on results described in this paper is currently in preparation and will be presented elsewhere.

While microlensing promises to help with studies of cosmological sources and intervening compact object populations, it may also contribute to uncertainty in statistics of multiple imaging, both by changing the level of magnification bias in gravitational lens surveys and by causing variation in the flux ratios of multiply imaged quasars. Many authors have noted that optically measured flux ratios cannot be used as reliable constraints for galaxy lens models due to microlensing of one or more images (e.g. Schneider et al.~1988; Kochanek~1991; Witt, Mao \& Schechter~1995; Mediavilla et al.~1998). Indeed, variation between optical and radio flux ratios (the radio should not be subject to microlensing) has been observed in several lensed quasars (e.g. Falco et al.~1996; Schechter \& Moore~1993; Katz \& Hewitt~1993). The effect of microlensing on magnification bias and flux ratios was discussed by Bartelmann \& Schneider~(1990). They suggested that the effect could be severe, particularly on the flux ratios. However, their model was simplistic. In particular they assumed a point source, and analytic approximations for magnification patterns. On the other hand, with the exception of Koopmans \& Wambsganss~(2000) and Loeb \& Perna~(1997) (who used results of Bartelmann \& Schneider~1990), this is to our knowledge the only previous paper to consider microlensing statistics based on the distribution of optical depth and shear for lines of sight to background cosmological sources.

The growing base of different observations for an increasing number of systems motivates us to explore the a-priori probabilities for the observation of microlensing phenomena, as well as the circumstances under which that microlensing will be observed. The work is performed with views towards determining the types of microlensing observations that will be plausible in the future, and towards testing our understanding of the populations involved in much the same spirit as investigations of macrolensing probabilities (e.g. Turner, Ostriker \& Gott~1984; Turner~1990; Kochanek~1996). We begin in Sec.~\ref{jointprob} by describing our model which is based on that described in Paper I, as well as the method for computing the probability distribution for the microlensing optical depth and shear. In Sec.~\ref{MLprob} we discuss magnification distributions for lensed images including microlensing, and the \emph{cross-section} for microlensing variability. Furthermore, we discuss the quasar microlensing variability statistics expected from two different classes of observational survey. We then discuss the effect of microlensing on the magnification bias for gravitational lens surveys and the distribution of image flux ratios for multiply imaged quasars. Finally in Sec.~\ref{GRB} we compute microlensing cross-sections for GRB afterglows. Macrolensing by galaxies results in two classes of images. Throughout the paper we refer to images of sources that are macrolensed as \emph{multiply imaged}, and those that are not as \emph{singly imaged}. We assume a flat cosmology having $\Omega=0.3$, $\Lambda=0.7$ and $H_0=65\,km\,sec^{-1}\,Mpc^{-1}$. Numerical tables of distributions presented in this paper will be made available upon request.

\section{The Joint Cross-Section for $\kappa$ and $\gamma$}
\label{jointprob}

This section describes the calculation of the joint probability distribution for the microlensing optical depth $\kappa$ and shear $\gamma$ near images of randomly distributed cosmological sources. This probability distribution will be used in calculation of magnification distributions and microlensing statistics in subsequent sections. Paper~I presented the distribution of $\kappa$ due to the normal stellar populations of galaxies for quasar images with random source positions. The model used is summarized below, however we refer the reader to Paper I for further details. 

A constant co-moving number-density of galaxies was assumed. The total mass-distributions of these galaxies were described by isothermal spheres having a central velocity dispersion $\sigma_{DM}=f\sigma$ (where $\sigma$ is the observed central velocity dispersion\footnote{$f$ is the correction factor between the velocity dispersions of the luminous and dark matter ($f=\sqrt{\frac{3}{2}}$ obtained for the simplest dynamical models was introduced by Turner, Ostriker \& Gott~1984). Kochanek~(1996, and references therein) advocates a factor close to unity from dynamical modeling of nearby early-type galaxies and from the distribution of observed macrolens image separations.} and $f\sim1$). The stellar populations embedded in these galaxies form the microlens population. In elliptical/S0 galaxies, stars were distributed with de-Vaucouleurs profiles (mass-to-light ratios were assumed constant in radius and red-shift), and in spiral galaxies with de-Vaucouleurs bulges and Kusmin (1956) discs. The distribution of velocity dispersions was described by the combination of a Schechter function $d\Phi/d\sigma$ and the Faber-Jackson~(1976) relation. The characteristic velocity dispersion of an $L_\star$ galaxy was taken to be $\sigma_\star=220{\rm \,km\,sec^{-1}}$.

In this work we follow a similar prescription. However, following the result of paper I that spiral galaxies contribute $\la10\%$ to the microlensing rate, we consider only elliptical/S0 galaxies in our microlensing calculations. Furthermore, we allow the microlens mass-density to evolve with redshift in proportion to the cumulative star-formation history\footnote{The results are quite insensitive to the star-formation history since most microlensing is due to galaxies at redshifts below 1. The exception is in the rate of microlensing of singly imaged quasars by stars in galaxies at higher redshifts.}. A simple model which has constant star-formation from $z=10$ till $z=1$ and then a rate proportional to $(1+z)^{3}$ until the present day was used (e.g. Hogg~1999 and references therein; Nagamine, Cen \& Ostriker~2000). The mass-to-light ratios of the galaxies were then normalized so that the elliptical plus spiral galaxy populations contain the cosmological density in stars at redshift zero [$\Omega_\star=0.005$ (Fukugita, Hogan \& Peebles~1998)]. The model of a given elliptical galaxy with redshift $z_s$ and central velocity dispersion $\sigma$ has an overall convergence at radius $\xi$ of
\begin{equation}
\kappa_{SIS} = \frac{1}{\Sigma_{crit}}\frac{\sigma^2}{2G}\frac{1}{\xi},
\end{equation}
and an average local convergence in stars of
\begin{equation}
\kappa_{*}= \Upsilon \frac{\Sigma_0}{\Sigma_{crit}} 10^{3.33\left(1-\left(\frac{\xi}{R_0}\right)^{\frac{1}{4}}\right)}.
\end{equation}
where 
\begin{equation}
\Sigma_{crit} = \frac{c^2}{4\pi G}\frac{D_s}{D_d D_{ds}}
\end{equation}
is the critical density for lensing, $R_0$ and $\Sigma_0$ (the characteristic radius and central density) are functions of $\sigma$ (see Paper I, after Djorgovski \& Davis~1987), and $\Upsilon$ is the mass-to-light ratio of the stellar population. Assuming the dark halo to be smoothly distributed, this results in a smooth component of convergence given by $\kappa_c = \kappa_{SIS}-\kappa_\star$. The shear is $\gamma_{*,c} = \kappa_{SIS}=\kappa_\star+\kappa_c$. In the above expressions $D_d$ and $D_s$ are the angular diameter distances of the lens and source at redshifts of $z_d$ and $z_s$, and $D_{ds}$ is the angular diameter distance from the lens to the source. Galaxies with small velocity dispersions can have $\kappa_\star>\kappa_{SIS}$ at very small radii using this definition, however we constrain $\kappa_\star\le\kappa_{SIS}$ for all $\xi$, thus keeping $\kappa_c\ge0$. 

Microlensing statistics are described by a combination of the stellar mass-density and the perturbing effects of a continuous component of convergence as well as shear from the galactic mass distribution (See Webster et al.~1992 for a detailed description of the dependencies in the case of Q2237+0305). Therefore three parameters ($\kappa_\star$, $\kappa_c$, $\gamma_{*c}$) govern the microlensing statistics of any one microlensed image. However, to reduce computation, we use a parameter transformation (Paczynski~1986; Kayser, Refsdal \& Stabell~1986; Schneider \& Weiss~1987) which allows the microlensing statistics to be described by an equivalent model having no continuous component of convergence, and thus only two microlensing parameters ($\kappa, \gamma$). Note that this transformation describes a degeneracy, and that the dependence due to $\kappa_c$ is merely shifted to other physical parameters. The lens equation may be written in component form 
\begin{eqnarray}
\nonumber
&&y_{1,*c}=(1-\kappa_c-\kappa_\star-\gamma_{c*})x_{1,*c}\\
&&y_{2,*c}=(1-\kappa_c-\kappa_\star+\gamma_{c*})x_{2,*c},
\end{eqnarray}
where $x=\xi/\xi_{o,*c}$ and $y=\eta/\eta_{o,*c}$ are the image and source positions in units of the microlens Einstein radius in the lens and source plane.
Given point-source magnification $\mu_{*c}$ and source size $S_{*c}$, the transformations 
\begin{eqnarray}
\label{defns}
\nonumber
&&\kappa=\frac{\kappa_\star}{|1-\kappa_c|}\\
\nonumber
&&\gamma=\frac{\gamma_{*c}}{|1-\kappa_c|}\\
\nonumber
&&\mu = \mu_{*c}\times(1-\kappa_c)^2\\
\nonumber
&&S = S_{*c}/\sqrt{|1-\kappa_c|}\\
\nonumber
&&\eta_o = \eta_{o,*c}\times\sqrt{|1-\kappa_c|}\\
\nonumber
&&\xi_o = \xi_{o,*c}/\sqrt{|1-\kappa_c|}\\
\nonumber
&&y = y_{*c}/\sqrt{|1-\kappa_c|}\\
&&x = x_{*c}\times\sqrt{|1-\kappa_c|}
\end{eqnarray}
yield an equivalent lens equation with identical microlensing properties, but no continuous component of convergence, hence
\begin{eqnarray}
\nonumber
&&y_1=(1-\kappa-\gamma)x_1\\
&&y_2=(1-\kappa+\gamma)x_2.
\end{eqnarray}
Below we find the differential joint cross-section $\frac{d^2\tau}{d\kappa d\gamma}$ for $\kappa$ and $\gamma$, as well as probabilities for the sets of parameters $\{\kappa$, $\gamma$, $\kappa_c$, $z_d$\}. The parameters $\kappa$, $\gamma$ and $\kappa_c$ are not independent quantities and the following procedure is used to find $\frac{d^2\tau}{d\kappa d\gamma}$. First, the probability of an image subject to a microlensing optical depth between $\kappa$ and $\kappa+\Delta\kappa$ at a redshift between $z_d$ and $z_d+\Delta z_d$ is computed (see paper I for details) by taking the derivative with respect to $\kappa$ of the differential cross-section for an image being subject to a microlensing optical depth larger than $\kappa$
\begin{equation} 
\frac{d^2\tau}{d\kappa dz_d}=\frac{d}{d\kappa}\frac{d\tau}{dz_d}(>\kappa).
\end{equation}
The relation $\xi_{\kappa}=\xi(\kappa)$ is multi-valued, having $2N_{pairs}=2$ or 4 solutions. 
At each of the $2N_{pairs}$ solutions of $\xi_{\kappa}=\xi(\kappa)$ we find the constants $C1^i$ and $C2^i$ such that
\begin{eqnarray}
\nonumber
&&C1^i \propto \eta \left.\frac{d\xi}{d\eta}\right|_{\xi=\xi_{\kappa_1^i}}\left.\frac{d\kappa}{d\xi}\right|_{\xi=\xi_{\kappa_1^i}},\\
&&C2^i \propto \eta \left.\frac{d\xi}{d\eta}\right|_{\xi=\xi_{\kappa_1^i}}\left.\frac{d\kappa}{d\xi}\right|_{\xi=\xi_{\kappa_2^i}}
\end{eqnarray} 
and
\begin{equation}
\sum_{i=1}^{i=N_{pairs}}(C1^i + C2^i) = 1.
\end{equation}
We also find the values of microlensing shear ($\gamma$$_1^i$, $\gamma$$_2^i$), and the values of smooth matter density ($\kappa$$_{c,1}^i$, $\kappa$$_{c,2}^i$). This procedure is repeated over a logarithmic grid of values for $\kappa$, $\sigma$ and $z_d$, with spacings $\Delta\kappa$, $\Delta\sigma$ and $\Delta z_d$. The resulting sets of values, $\{\kappa$, $\kappa$$_{c,1}^i$, $\gamma$$_1^i$, $z_d\}$ and $\{\kappa$, $\kappa$$_{c,2}^i$, $\gamma$$_2^i$, $z_d\}$ have corresponding probabilities 
\begin{eqnarray}
\nonumber
p(\{\kappa,\gamma_1^i,\kappa_{c,1}^i,z_d\})\Delta\kappa\Delta z_d=C1^i\frac{d^2\tau}{d\kappa dz_d}|_{\xi=\xi_{\kappa_1^i}}\frac{d\Phi}{d\sigma}\Delta\sigma \Delta\kappa\Delta z_d\\
\nonumber\mbox{ and }\\
p(\{\kappa,\gamma_2^i,\kappa_{c,2}^i,z\})\Delta\kappa\Delta z_d=C2^i\frac{d^2\tau}{d\kappa dz_d}|_{\xi=\xi_{\kappa_2^i}}\frac{d\Phi}{d\sigma}\Delta\sigma\Delta\kappa\Delta z_d.
\end{eqnarray}
These probabilities are binned in $\gamma$ and integrated over $z_d$, to find the the differential joint cross-section 
\begin{eqnarray}
\nonumber
&&\frac{d^2\tau}{d\kappa d\gamma} = \int_0^{z_s}dz_d\int_0^{\infty}d\gamma' H(\gamma,\Delta\gamma)\frac{p(\{\kappa,\gamma',\kappa_{c},z\})}{\Delta \gamma}\\
\nonumber
&&\mbox{where}\\
&&H(\gamma,\Delta\gamma)=\left\{\begin{array}{l}1\hspace{5mm}\gamma-\frac{\Delta\gamma}{2}<\gamma'<\gamma+\frac{\Delta\gamma}{2}\\
0\hspace{5mm}\mbox{otherwise}
\end{array}\right.
\end{eqnarray}
of observing an image subject to a microlensing optical depth between $\kappa$ and $\kappa+\Delta\kappa$, and microlensing shear between $\gamma$ and $\gamma+\Delta\gamma$. For impact parameters smaller than the galaxy Einstein radius, the SIS lens model produces 2 images. For multiply imaged sources the values of $\kappa$, $\gamma$ and $\kappa_c$ for the second image are kept in the set in addition to those of the primary image. The parameters for the second image are easily computed by noting that the image separation for an SIS is always $\Delta x=2$. These additional parameters facilitate calculation of the distribution of total magnification for pairs of multiple images (Sec.~\ref{magdist}). 

\begin{figure*}[hptb]
\epsscale{1.}
\plotone{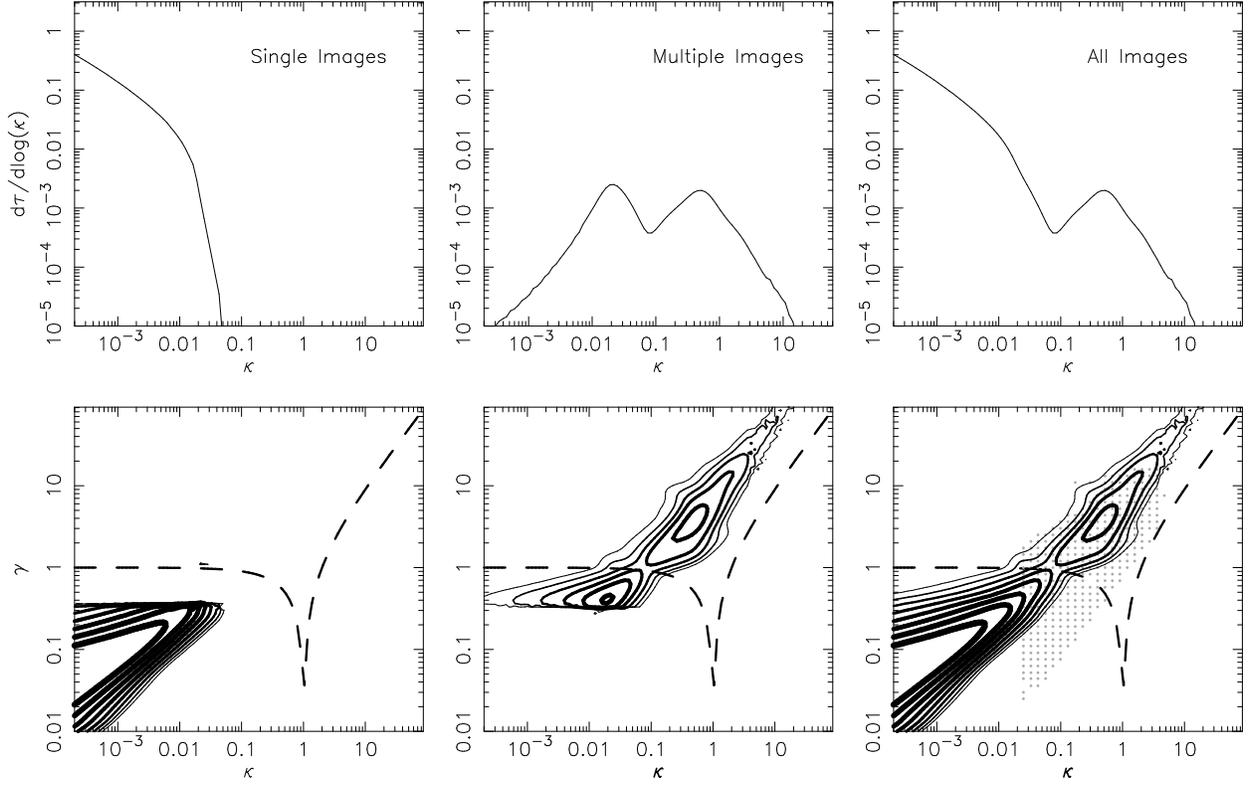}
\caption{Lower Rows: Contour plots (with contours spaced by multiples of $\sqrt{10}$) for the joint differential cross-section of $\kappa$ and $\gamma$. The dashed lines show the condition for infinite magnification $1-\kappa=\pm\gamma$, and the dots in the right panel the parameters of the computed magnification maps. Upper panels: Differential probability for $\kappa$. The left, center and right panels show probabilities for single, multiple and all images respectively, the source redshift was $z_s=3$.}
\label{fig1}
\end{figure*}

The lower rows of Fig.~\ref{fig1} show contour plots of the differential joint cross-section $\frac{d^2\tau}{d\kappa d\gamma}$ for sources that are singly imaged by the galaxy (left), multiply imaged by the galaxy (center), and for all images (right). The case shown has a source redshift of $z_s=3$ and the contours are spaced by factors of $\sqrt{10}$. Several features of the distributions warrant explanation: 

\noindent 1) $\kappa>\gamma$ for all $\kappa$, as can be easily seen from their definitions (Eqn.~\ref{defns}). 

\noindent 2) There is a line of demarcation between regions of $\kappa-\gamma$ parameter space accessible to single and multiple images. Multiple images formed by an isothermal sphere have $\kappa_{SIS}=\kappa_\star+\kappa_c=\gamma_{SIS}=\gamma_{*c}>1/4$. This demarcation line is therefore parameterized by
\begin{eqnarray}
\nonumber
&&\kappa = \frac{\kappa_\star}{|\kappa_\star+3/4|}\hspace{5mm}\mbox{and}\\
&&\gamma = \frac{1/4}{|\kappa_\star+3/4|}.
\end{eqnarray} 
Given positive $\kappa_\star$, this yields $\gamma=\frac{1-\kappa}{3}$ for $\kappa<1$ and $\gamma<1/3$. 

\noindent 3) An SIS has $\kappa_{SIS}=\kappa_\star+\kappa_c=\gamma_{SIS}=\gamma_{*c}=1/2$ at the Einstein radius. The relation between $\kappa$ and $\gamma$ at the Einstein radius is therefore parameterized by
\begin{eqnarray}
\nonumber
&&\kappa = \frac{\kappa_\star}{|\kappa_\star+1/2|}\hspace{5mm}\mbox{and}\\
&&\gamma = \frac{1/2}{|\kappa_\star+1/2|},
\end{eqnarray} 
which results in the relation $\gamma = 1-\kappa$. This is the condition of formally infinite magnification required at the Einstein radius. The line $1/\mu=|(1-\kappa)^2-\gamma^2|=0$ is shown as the dashed line in the lower panels of Fig.~\ref{fig1}. Note that this line runs through a valley in the probability density, which is the result of depletion of images near the galaxies' Einstein radii.
The upper panels of Fig.~\ref{fig1} show $\frac{d\tau}{d\kappa}$ obtained by integrating $\frac{d^2\tau}{d\kappa d\gamma}$ over $\gamma$. The plots clearly show the depletion near the Einstein Radius.

\section{The Probability for Quasar Microlensing}
\label{MLprob}

In this section we calculate microlensing statistics by combining probabilities for different combinations of microlensing parameters $p(\{\kappa,\gamma,\kappa_{c},z_d\})\Delta\kappa\Delta z_d$ with results from numerical microlensing simulations. A large number (556) of magnification patters were computed over the region of interest defined by $\frac{d^2\tau}{d\kappa d\gamma}$. The computation of these magnification patterns was performed using the \emph{microlens} ray-tracing program, generously provided by Joachim Wambsganss. The values of $\kappa$ and $\gamma$ for which magnification patterns were computed are marked (light dots) in the lower right panel of Fig.~\ref{fig1}. Each magnification pattern was 25 microlens Einstein radii on a side. One magnification pattern of this size does not adequately describe the microlensing statistics for the corresponding set of microlensing parameters (Seitz, Wambsganss \& Schneider~1994), and as a result probabilities calculated for a single set of parameters will not be accurate. On the other hand, our calculations of microlensing probabilities average over a large number of patterns having different microlensing parameters, so that simulation variance will average out. Furthermore, we normalize each magnification pattern by the corresponding theoretical magnification ($\mu_{th}=|(1-\kappa)^2-\gamma^2|^{-1}$). 
Magnification patterns having $\kappa<0.025$ were not computed, and statistics of microlensing for ($\kappa,\gamma)$ where $\kappa<0.025$ were found by multiplying the probabilities computed using the (0.025, $\gamma$) magnification pattern by $(1-e^{-\kappa})/(1-e^{-0.025})$. We computed 10 realizations of each of the 20 magnification patterns at $\kappa=0.025$, so that statistics for a given shear with  $\kappa<0.025$ would not be heavily biased by any single pattern.

In paper I, the Poisson probability that a source will lie inside the Einstein ring of at least one microlens was used to approximate the conditional probability distribution for the microlensing optical depth near lines of sight to microlensed quasars
\begin{equation}
\label{probML}
\frac{d^2\tau_{ML}}{d\kappa dz}=(1-e^{-\kappa})\frac{d^2\tau}{d\kappa dz}.
\end{equation}
This approximation has several shortcomings. Firstly, the contribution of $\gamma$ to microlensing statistics is ignored. Secondly, the probability of microlensing at $\kappa\sim1$ is assumed to be $\sim1$, while simulations at $\kappa\pm\gamma\sim1$ show that the size of light-curve fluctuations tends to zero (Deguchi \& Watson 1987; Seitz, Wambsganss \& Schneider~1994). Thirdly, the source size is not considered, even though a larger source size results in longer timescales and reduced event amplitudes for microlensing events (Wambsganss, Paczynski \& Katz~1989). In addition to these shortcomings, the quantity computed by Eqn.~\ref{probML}, namely the fraction of sources microlensed at any one time is not necessarily the quantity of interest. Rather it is more useful to know the magnification distribution resulting from microlensing, and the likelihood of variability above some threshold level. In the following subsections we consider these in turn.

In what follows, the source quasar is assumed to have a radius of $S_{*,c}=10^{15}$cm (typical scale for an accretion disc; found for Q2237+0305 by Wambsganss, Paczynski \& Schneider~1990 and Q0957+561 by Refsdal, Stabell, Pelt \& Schild~2000) and to have a uniform top-hat profile. We note that the size of the emission region for a fixed observational band will decrease with source redshift (for a thermal accretion disc) since we are observing intrinsically higher frequencies and therefore brightness temperatures. Since $\frac{d^2\tau}{d\gamma d\kappa}$ has been computed for the effective microlensing optical depth and shear defined by Eqn.~\ref{defns}, the physical value of source size must be adjusted accordingly for each value of $\kappa_c$. The mass-spectrum is demonstrably unimportant (e.g. Witt, Kayser \& Refsdal~1993), and the microlenses are assumed to have a single mass of $0.1M_{\odot}$.

\subsection{Magnification Distributions}
\label{magdist}

\begin{figure*}[hptb]
\epsscale{1.}
\plotone{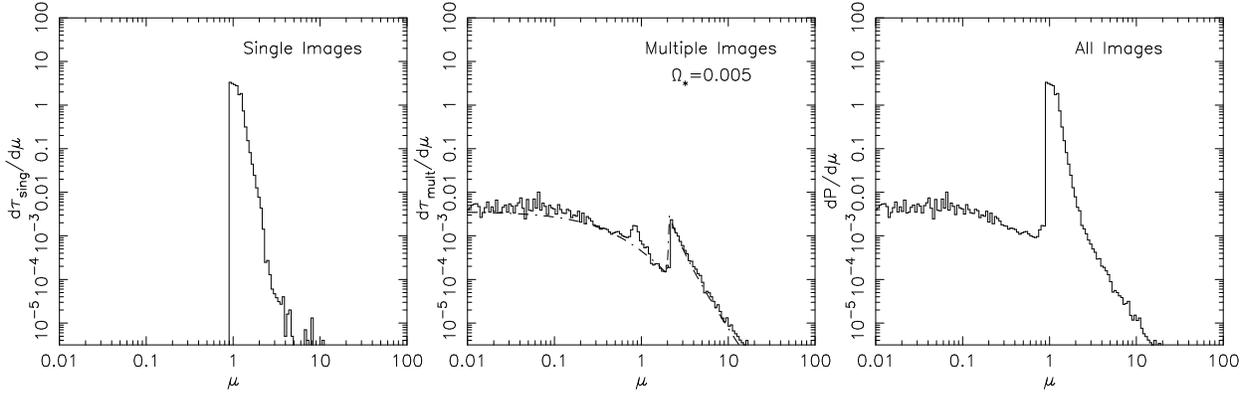}
\caption{Magnification distributions for quasar images. The left, center and right panels show distributions for single, multiple and all images respectively. The dot-dashed lines in the central panel show the distribution for a smooth isothermal sphere and the source redshift was $z_s=3$.}
\label{fig2}
\end{figure*}

Recall that we find a probability $p(\{\kappa,\gamma,\kappa_c,z_d\})\Delta\kappa\Delta z_d$ for each set of values $\{\kappa,\gamma,\kappa_c,z_d\}$. To find the magnification distribution due to galaxies containing populations of stars, these probabilities are convolved with the magnification distribution computed for the corresponding sets of microlensing parameters. A coarse magnification distribution $\frac{dP}{d\mu}(\kappa,\gamma,\kappa_c,z_d)$ was obtained for each parameter set by binning 100 magnifications generated from  the magnification map having microlensing parameters closest to those of the current set (we also normalize the mean of the magnification pattern by the theoretical mean corresponding to the current set of parameters). For each parameter set the source size was determined in units of effective microlens Einstein radius ($S/\eta_o=S_{*c}/\eta_{o,*c}\times|1-\kappa_c|$). The physical magnifications were then determined from the scaled magnifications $\mu_{*c} = \mu/(1-\kappa_c)^2$ which were computed by convolving the source profile with the magnification map. The resulting distributions were then integrated over $z_d$ and $\kappa$. 
\begin{equation}
\frac{d\tau}{d\mu} = \int_{10^{-4}}^{\infty}d\kappa\int_0^{z_s}dz_d p(\{\kappa,\gamma,\kappa_{c},z_d\})\frac{dP}{d\mu}(\kappa,\gamma,\kappa_c,z_d)
\end{equation}
The procedure misses lines of sight having $\kappa$ smaller than $10^{-4}$ (the smallest value considered). However we know that the single image distribution [$\frac{d\tau_{sing}}{d\mu}$] must be normalized to $1-\tau_{mult}$ (where $\tau_{mult}$ is the cross-section for multiple imaging), and that its mean must be $(1-4\tau_{mult})/(1-\tau_{mult})$ so that the average of the magnification distribution [$\frac{dP}{d\mu}=\frac{d\tau_{mult}}{d\mu}+\frac{d\tau_{sing}}{d\mu}$] for all quasars is unity. These conditions were fulfilled by adding probability smoothly to the bins between $\mu=0.9$ and 1.1 [we are not concerned with the details of the distribution near $\mu=1$ which must be computed from n-body simulations (e.g. Barber, Thomas, Couchman \& Fluke~2000)]. Inaccuracies can arise from the integration over the singularity in magnification as a function of $\kappa$ and $\gamma$ using a finite grid. The multiple image distribution was corrected for this effect by multiplying by the analytic SIS distribution over the numerical distribution assuming a smooth mass distribution (i.e. no microlensing). Magnification distributions were computed for $z_s=1$, 2, 3 and 4. Example distributions for a source at $z_s=3$, are plotted in Fig.~\ref{fig2} for single ($\frac{d\tau_{sing}}{d\mu}$, left), multiple ($\frac{d\tau_{mult}}{d\mu}$, center) and all ($\frac{dP}{d\mu}=\frac{d\tau_{sing}}{d\mu}+\frac{d\tau_{mult}}{d\mu}$, right) images. The distribution for multiple images includes individual values for both the bright and faint macro-images. Also shown for comparison in the central panels is the analytic distribution for a smooth SIS. The figure shows that microlensing results in a spread of both the single and the multiple image magnification distributions. Of particular note in the multiple image distribution is the small excess of large magnifications over the smooth SIS level. In the distribution for single images, we see that the effect of microlensing is to create a significant non-zero probability for magnifications with values greater than 2, which are not formed by the SIS.

\begin{figure*}[hptb]
\epsscale{.4}
\plotone{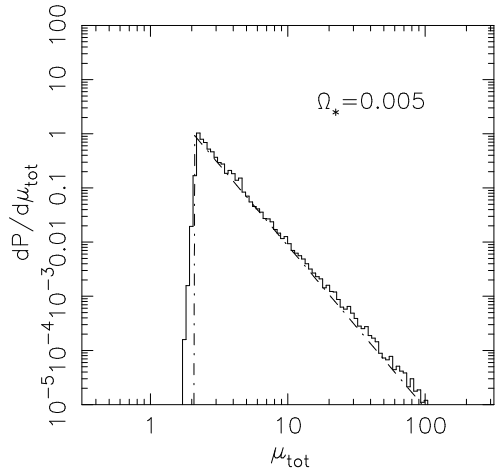}
\caption{Magnification distributions for the sum of images in multiply imaged quasars. The dot-dashed lines show the distribution for a smooth isothermal sphere and the source redshift was $z_s=3$.}
\label{fig3}
\end{figure*}

The distribution ($\frac{dP}{d\mu_{tot}}$) of total magnifications ($\mu_{tot}$, the sum of the magnifications for the bright and faint images) for multiply imaged quasars was computed using an analogous procedure. However instead of computing the distribution of magnifications for a single magnification pattern, the distribution was computed for the sum of magnifications computed from pairs of magnification patterns. These magnification patterns had microlensing parameters corresponding to those recorded for both images in the sets of parameters described in Sec~\ref{jointprob}. The resulting distribution is shown in Fig.~\ref{fig3} for a source at $z_s=3$. The analytic distribution for a smooth SIS is again shown for comparison. We find an excess of large magnifications and a non-zero probability for $\mu_{tot}<2$, the minimum value for the total magnification of a smooth SIS. In Sec.~\ref{bias} we will use this distribution to compute the effect of microlensing on the magnification bias for multiple imaging in optical gravitational lens surveys.

\subsection{The Probability of Quasar Microlensing Variability}

In this subsection we compute the probability that the magnitude of a quasar will vary by more than $\Delta m$ due to microlensing during a 10 year period. For this calculation we assume that the velocity components $v_{x}$, and $v_{y}$ of the galaxy ($v_d$), source ($v_s$) and observer ($v_o$) are Gaussian distributed with $\sigma_v=400km\,sec^{-1}$. From Kayser, Refsdal \& Stabell (1986) we compute the effective source plane transverse velocity
\begin{eqnarray}
\nonumber
&&v_{eff,*c}=\sqrt{v_{eff,*c,x}^2+v_{eff,*c,y}^2},\hspace{5mm}\mbox{where}\\
\nonumber
&&v_{eff,*c,x} = \frac{v_{o,x}}{1+z_d}\frac{D_{ds}}{D_d}  +  \frac{v_{d,x}}{1+z_d}\frac{D_s}{D_d}   +  \frac{v_{s,x}}{1+z_s}\hspace{5mm}\mbox{and}\\
&&v_{eff,*c,y} = \frac{v_{o,y}}{1+z_d}\frac{D_{ds}}{D_d}  +  \frac{v_{d,y}}{1+z_d}\frac{D_s}{D_d}   +  \frac{v_{s,y}}{1+z_s}.
\end{eqnarray}
Since the probabilities $p(\{\kappa,\gamma,\kappa_{c},z_d\})\Delta\kappa\Delta z_d$ have been computed in terms of the effective microlensing optical depth and shear defined by Eqn.~\ref{defns}, the physical value of transverse velocity must be adjusted accordingly (Eqn.~\ref{defns}) for each value of $\kappa_c$
\begin{equation}
v_{eff} = v_{eff,*c}/\sqrt{|1-\kappa_c|}.
\end{equation}

\subsubsection{microlensing variability cross-sections}
\label{cross-sections}

\begin{figure*}[hptb]
\epsscale{1.}
\plotone{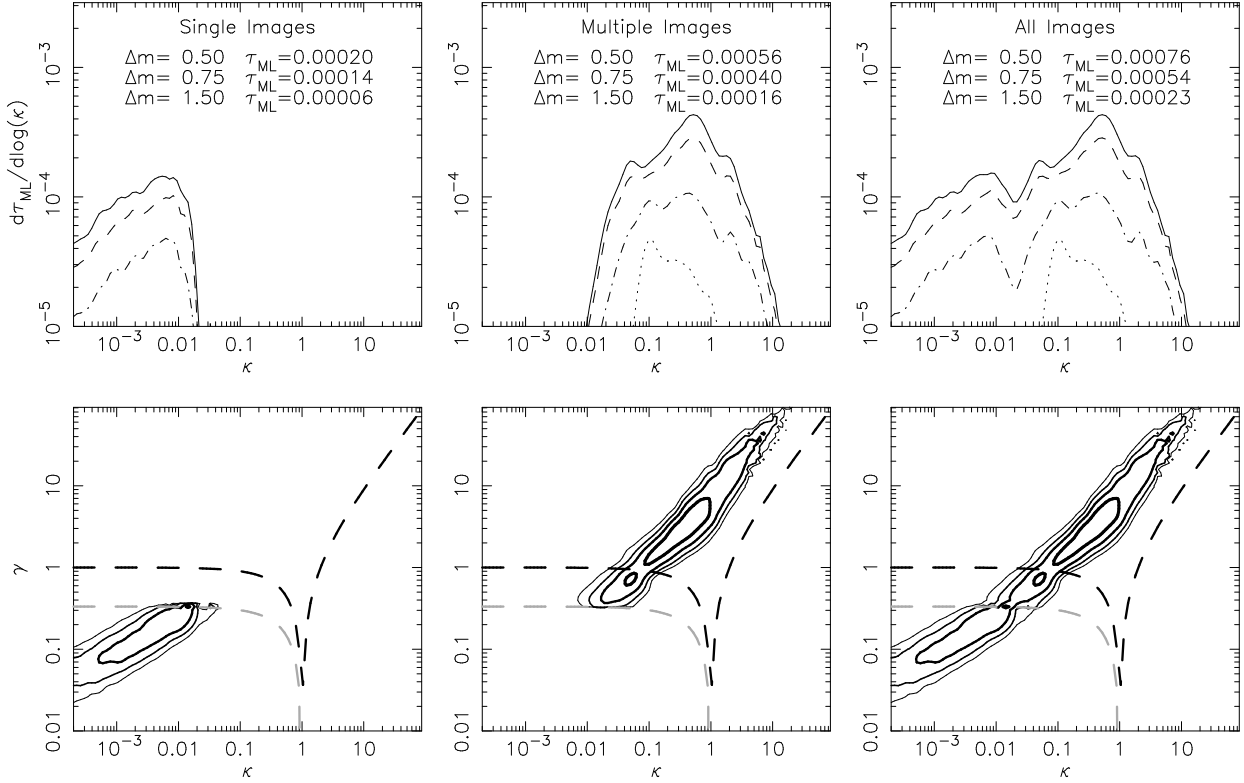}
\caption{Lower Rows: Contour plots (with contours spaced by multiples of $\sqrt{10}$) for the joint differential cross-section for $\kappa$ and $\gamma$ near the lines of sight to microlensed ($\Delta m>0.5$) quasars. The dark dashed lines show the condition for infinite magnification $1-\kappa=\pm\gamma$, and the light dashed lines separate regions of single and multiple images. Upper panels: Differential probability for $\kappa$ near the lines of sight to microlensed quasars. The solid, dashed, dot-dashed and dotted lines correspond to $\Delta m>0.5$, 0.75, 1.5 and 2.5. The left, center and right panels show probabilities for single, multiple and all images respectively, the source redshift was $z_s=3$}
\label{fig4}
\end{figure*}

Again recall that we find a probability $p(\{\kappa,\gamma,\kappa_{c},z_d\})\Delta\kappa\Delta z_d$ for each set of values $\{\kappa,\gamma,\kappa_c,z_d\}$. To find the conditional joint differential cross-section for values of $\kappa$ and $\gamma$ near lines of sight to microlensed images, these probabilities were multiplied by the probability $f$ that the source will be microlensed. In this case we take $f(\Delta m|\kappa,\gamma,\kappa_c,z_d)$ to be the fraction of light-curves $m(t)$ with 10 year monitoring periods that vary by more than $\Delta m$ (note, in paper I $f$ was taken to be $1-e^{-\kappa}$, Eqn.~\ref{probML}). We computed $f(\Delta m|\kappa,\gamma,\kappa_c,z_d)$ for each parameter set from 100 light-curves generated using the magnification map having microlensing parameters closest to those of the current set (we also normalize the mean of the magnification pattern to that of the current set of parameters). The light-curves had directions perpendicular, parallel and at 45 degrees to the shear with weightings of 0.25, 0.25 and 0.5 respectively which approximates random directions for the transverse velocity. For each parameter set the transverse velocity in units of Einstein radii per second is $\frac{v_{eff}}{\eta_{o}} = \frac{v_{eff,*c}}{\eta_{o,*c}}\frac{1}{|1-\kappa_c|}$. The true magnification is related to the magnification computed from the integral over source size by $\mu_{*c} = \mu/(1-\kappa_c)^2$, and the resulting light-curve is $m(t)=-2.51\log(\mu_{*c}) + const$. Values of $f(\Delta m|\kappa,\gamma,\kappa_c,z_d)\times p(\{\kappa,\gamma,\kappa_c,z_d\})$ were then binned in $\gamma$ and integrated over $z_d$ as before to yield the conditional differential joint cross-section for values of $\kappa$ and $\gamma$ near lines of sight to microlensed images
\begin{equation}
\frac{d^2\tau_{ML}}{d\kappa d\gamma} = \int_0^{z_s}dz_d\int_0^{\infty}d\gamma' H(\gamma,\Delta\gamma)\frac{f(\Delta m|\kappa,\gamma,\kappa_c,z_d)\times p(\{\kappa,\gamma',\kappa_{c},z_d\})}{\Delta\gamma}.
\end{equation}
The lower panels of Fig.~\ref{fig4} show $\frac{d^2\tau_{ML}}{d\kappa d\gamma}$ for the case of a source at $z_s=3$ and $\Delta m=0.5$ for singly imaged (left panel), multiply imaged (considering each image separately, center panel) and all (right panel) sources. The distribution shows many of the features seen in Fig.~\ref{fig1}, namely depletion near the Einstein Radius, demarcation between regions having singly and multiply (by the galaxy) imaged sources (light dashed line), and $\kappa<\gamma$ for all $\kappa$. However, the distribution is suppressed in the regions of both low and high $\kappa$. The upper panels of Fig.~\ref{fig4} show $\frac{d\tau_{ML}}{d\kappa}$ obtained by integrating $\frac{d^2\tau_{ML}}{d\kappa d\gamma}$ over $\gamma$. Distributions are plotted for $\Delta m=0.5$, 1.0 and 1.5 magnitudes. The integrals of these distributions $\tau_{ML}(\Delta m,z_s)$, which we term the \emph{microlensing cross-section}, i.e. the fraction of sources that vary by more than $\Delta m$ per decade due to microlensing are also given. 

\begin{figure*}[hptb]
\epsscale{1.}
\plotone{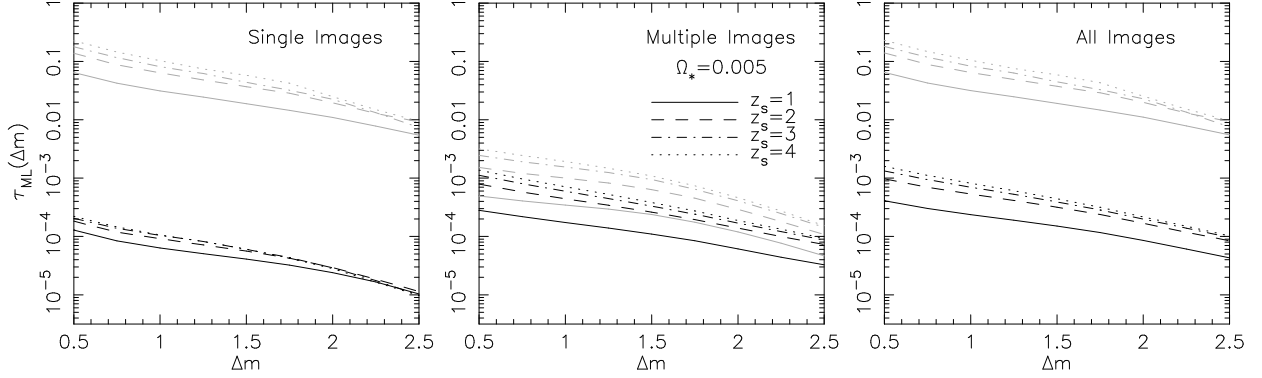}
\caption{Plots of the quasar microlensing cross-section $\tau_{ML}$ verses $\Delta M$. The left, center and right panels show values for single, multiple, and all images respectively. The dark lines correspond to microlensing by stars, while the light lines assume dark matter to be in the form of compact objects. Values are shown for $z_s=1$, 2, 3 and 4.}
\label{fig5}
\end{figure*}

The microlensing cross-section $\tau_{ML}(\Delta m,z_s)$ is plotted as a function of $\Delta m$ in Fig.~\ref{fig5} for $z_s=1$, 2, 3 and 4. The left, central and right panels show values for single quasar images, multiple quasar images (considering both images separately) and all images respectively. For example, $\sim1$ in 1000 quasar images at $z_s=3$ vary by more than $\Delta m=0.5$ magnitudes due to microlensing. the majority of this microlensing occurs in sources that are also multiply imaged by the galaxy. As expected, large amplitude variability is rarer than low amplitude variability, and there is a 10 fold decrease in rate between $\Delta m=0.5$ and $\Delta m=2.5$ magnitudes in all examples. Microlensed variability at all levels is more likely at higher source redshift. 

Microlensing cross-sections were also computed under the assumption that the dark matter is entirely composed of compact objects. The results are shown as the grey lines in Fig.~\ref{fig5}. The SIS does not have finite mass, and so the surface mass distributions were truncated at an outer radius $\xi_{max}$ (proportional to $\sigma^2$ for each galaxy) such that the sum of the masses constitutes the cosmological density $\Omega$. The numerical procedure was checked by noting that the mean of the distribution $\frac{1}{\tau}\frac{d\tau}{d\kappa}$ should equal the result obtained for the optical depth of randomly distributed objects (Press \& Gunn~1973; Turner, Ostriker \& Gott~1984). Since much of the mass inside the galaxies Einstein ring is in stars, the microlensing cross-section in multiple images is quite insensitive to the addition of dark compact objects, giving an enhancement of only a factor of $\sim2-3$. On the other hand, while the microlensing cross-section due to stellar populations is very small for single images due to the rapid decline in the density of stars beyond the critical radius for multiple imaging, most of the dark matter mass lies beyond this radius. As a result, the inclusion of dark compact objects boosts the single image microlensing cross-section significantly to $\sim10\%$. This is expected since in the formalism of Press \& Gunn~(1973) we find that the optical depth is a few tenths of $\Omega$. We note that this is a very coarse approximation to lensing by the large scale dark-matter distribution, which is a topic beyond the scope of this paper. However, an unavoidable consequence of Fig.~\ref{fig5} is that comparison between variability of multiply imaged quasars, and quasars with lines of sight near galaxies will be a powerful and achievable method for detection or rejection of compact objects as dark matter candidates in the halos of galaxies.

The microlensing cross-sections described above assume a 1-d dispersion in the proper motions of galaxies of $v=400\,km\,sec^{-1}$, microlens masses of $\langle m\rangle = 0.1M_{\odot}$ and a source size of $S=10^{15}cm$. However we can discuss the results qualitatively for other choices for these parameters by noting that event peak amplitudes scale with $\langle m\rangle^{1/4}$ and $S^{-1/2}$, while event rates scale with $\langle m\rangle^{-1/2}$ and $v$ (Witt, Kayser \& Refsdal~1993). Using these relations we approximate the general results for the microlensing cross-sections shown in Figs.~\ref{fig5} and elsewhere in the paper for a 1-d proper motion velocity dispersion of $v'$, source size $S'$ and microlens mass $\langle m\rangle'$ as
\begin{eqnarray}
\nonumber
&&\tau'_{ML}(\Delta m') \sim \left(\frac{v'}{400\,km\,sec^{-1}}\right)\left(\frac{\langle m\rangle'}{0.1M_{\odot}}\right)^{-\frac{1}{2}}\tau_{ML}(\Delta m)\\
\nonumber
&&\mbox{where}\\
&&\Delta m \sim \Delta m' - 2.51\log_{10}\left[\left(\frac{\langle m\rangle'}{0.1M_{\odot}}\right)^{\frac{1}{4}}\left(\frac{S'}{10^{15}cm}\right)^{-\frac{1}{2}}\right].
\end{eqnarray}
As an example, for a source $S'=10S$ microlensed by microlenses of mass $\langle m\rangle'=0.01M_{\odot}$ we find that $\tau'_{ML}(\Delta m')\sim3.16\tau_{ML}(\Delta m=\Delta m'+1.88)$. For $\Delta m'=0.5$ magnitudes, this results in $\tau'_{ML}(\Delta m')\sim 4\times10^{-4}$ or half the value of $\tau_{ML}(\Delta m=0.5)$.

\subsubsection{the probability of microlens redshift}

\begin{figure*}[hptb]
\epsscale{1.}
\plotone{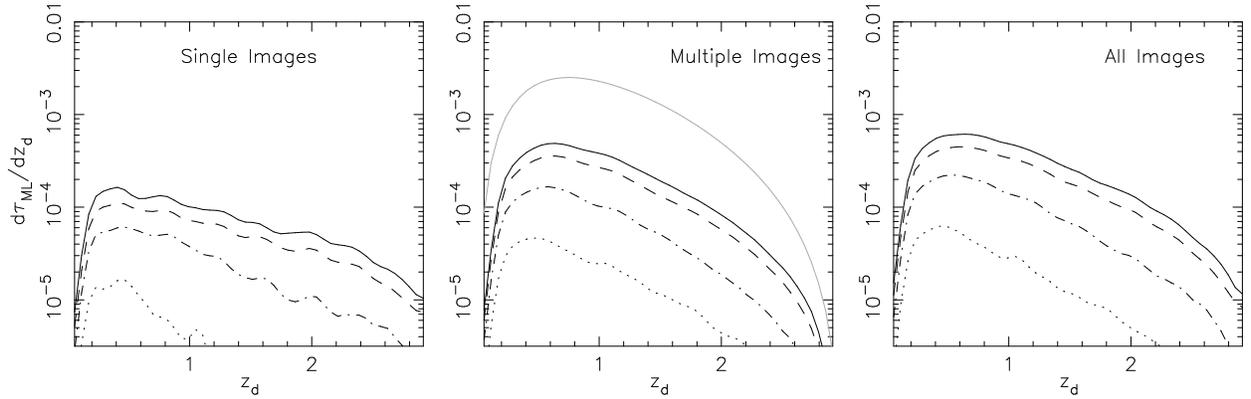}
\caption{Probability distributions for the redshift of galaxies responsible for microlensing of quasars. The left, center and right panels show values for single, multiple and all images respectively, and the solid, dashed, dot-dashed and dotted lines correspond to $\Delta m>0.5$, 0.75, 1.5 and 2.5. The light line in the central panel shows the distribution of macrolens redshifts. The source redshift was $z_s=3$.}
\label{fig6}
\end{figure*}

The probability distribution for the redshift of galaxies whose stellar populations result in the microlensing of back-ground sources was computed using an analogous approach. The resulting distributions $\frac{d\tau_{ML}}{dz_d}$ for the case of $z_s=3$ are plotted in Fig.~\ref{fig6} for singly imaged (left panel), multiply imaged (center panel) and all (right panel) sources. Distributions are plotted for $\Delta m=0.5$, 1.0, 1.5, and 2.5 magnitudes, and can be compared to the distribution of macrolens redshifts (light line). The distribution describes a low and narrow range of redshifts of microlensing galaxies that also produce multiple imaging. This peak is bounded from below by the low number of macro-lens galaxies at low redshift (e.g. Turner, Ostriker \& Gott 1984). From above the peak is bounded by two factors of $D_s/D_d$, one for the size of the source with respect to the microlens Einstein radius projected into the source plane (i.e. larger fluctuations for larger $D_s/D_d$), and one for the timescale which decreases in proportion to $D_s/D_d$ (resulting in an increase in microlensing rate). Furthermore, at higher redshifts the typical impact parameter for multiple images is larger with respect to the scale radius $R_o$ of the stellar distribution. Multiple images are therefore typically subject to lower values of $\kappa_\star$. As a result of these factors, the typical redshift of microlensing galaxies is lower than the typical macrolens galaxy redshift. While the microlensing cross-section at higher lens galaxy redshifts for singly imaged quasars is decreased due to the factors of $D_s/D_d$ already mentioned as well as from the reduced microlens population above $z_d\sim1$, the lower macrolensing cross-section for high-redshift lenses gives more single images lines of sight through regions of high $\kappa_\star$ in those galaxies. As a result $\frac{d\tau_{ML}}{dz_d}$ is broad for single images, in contrast to results for multiple images. Note however that the larger amplitude variability in single images is still only found for galaxies at low redshift since large amplitudes require a source that is small with respect to the caustic structure.

\subsubsection{observable microlensing statistics}

The results thus far have referred to probabilities for microlensing at fixed source red-shift, and have not included the effects of magnification bias. However, both the distribution of source redshifts, and the magnification bias need to be considered for a connection to be made with observations. In this subsection we find observed microlensing probabilities by combining differential cross-sections with the empirical double power-law quasar luminosity function $\Phi(L,z_s)$ of Pei~(1995) (based on data presented by Hartwick \& Schade~1990 and Warren, Hewett \& Osmer~1994) and the break luminosity evolution described by Madau, Haardt \& Rees~(1999). To convert the empirical luminosity function back to relative number counts above a limiting apparent B-magnitude ($m_B$) we convert $m_B$ to a limiting intrinsic luminosity $L_{lim}$ using the luminosity distance and a $k$-correction found from the procedure described in Pei~(1995) and M$\o$ller \& Jakobsen~(1991). The magnification due to gravitational lensing allows observations at a fixed limiting magnitude to reach further down the luminosity function and include more sources (e.g. Turner~1980). The resulting bias for the fraction of sources exhibiting the lensing phenomenon of interest often plays an important role in the analysis of lens samples. Magnification bias for a varying source requires careful treatment that depends in detail on the observed sample. In the remainder of this subsection we describe two different examples of the statistics of microlensing variability in hypothetical monitoring campaigns.  

\begin{figure*}[hptb]
\epsscale{1.}
\plotone{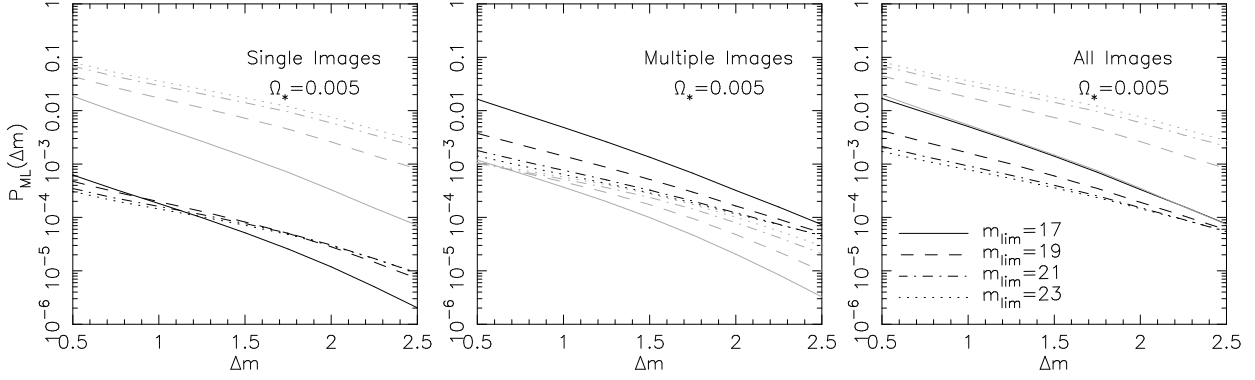}
\caption{Plots of the probability of microlensing $P_{ML}$ verses $\Delta M$ in a blind monitoring campaign. The left, center and right panels show values for single, multiple and all images respectively. The dark lines correspond to microlensing by stars, while the light lines assume dark matter to be in the form of compact objects. Values are shown for limiting B-magnitudes of $m_{lim}=17$, 19, 21 and 23.}
\label{fig7}
\end{figure*}

In the first example we compute the fraction of quasar images microlensed by more than $\Delta m$ per decade in a continuous blind survey to a fixed limiting magnitude $m_B$. This type of survey will be performed as part of the Sloan Digital Sky Survey (York et al.~2000) where a stripe of sky will be repeatedly scanned. Let $P_{sgle}(z_s)$ / $P_{mult}(z_s)$ be the fraction of quasar images that are microlensed and in single / multiple image systems (at source redshift $z_s$). These are given by
\begin{equation}
\label{MLfrac}
P(z_s,\Delta m,m_{lim}) = \frac{P'(z_s,\Delta m,m_{lim})F(z_s,\Delta m,m_{lim})}{P_{tot}'(z_s,\Delta m,m_{lim})} 
\end{equation}
where $P'_{tot}(z_s,\Delta m,m_{lim})=P_{sgle}'(z_s,\Delta m,m_{lim}) +P_{mult}'(z_s,\Delta m,m_{lim})$. The quantities $P'(z_s)=\tau(z_s) B(z_s)$ (where B is the magnification bias) are the fraction of images that are are in single or multiple images systems (images in multiple image systems are considered separately). The probability that an image is detected during the survey is the cross-section $\tau(z_s)$ multiplied by the bias calculated at the maximum light-curve magnification $\mu_{max}$. Since this maximum is subject to microlensing, the bias is given by the expression
\begin{eqnarray}
\label{B}
\nonumber
&&B(z_s,\Delta m,m_{lim})=\\
&&\frac{1}{\tau(z_s)}\int_{0}^{\infty} d\kappa\int_{0}^{z_s} dz_d p(\{\kappa,\gamma,\kappa_c,z_d\}|z_s)\left[\frac{\int_{0}^{\infty} d\mu_{max}\frac{dP}{d\mu_{max}}(\kappa,\gamma,z_d,z_s)N(>\frac{L_{lim}}{\mu_{max}},z_s)}{N(>L_{lim},z_s)}\right],
\end{eqnarray}
where $N(>L_{lim},z_s)=\int_{L_{lim}}^{\infty}dL\Phi(L,z_s)$.
The detection of variability larger than $\Delta m$ requires that the image be detectable when $\Delta m$ magnitudes fainter than the light-curve maximum. Therefore, while the bias for the detection of an image is computed using the magnification at the light-curve maximum, the bias for detection of a microlensed image should be calculated using a magnification of $\mu_{var}=\mu_{max}10^{-\Delta m/2.51}$. The parameter $F$ in Eqn.~\ref{MLfrac} is the fraction of images in single / multiple image systems that are microlensed by more than $\Delta m$. This is given by 
\begin{eqnarray}
\label{F}
\nonumber
&&\hspace{-30mm}F(z_s,\Delta m,m_{lim})=\\
\nonumber
&&\hspace{-15mm}\frac{1}{\tau(z_s)}\int_{0}^{\infty} d\kappa\int_{0}^{z_s} dz_d p(\{\kappa,\gamma,\kappa_c,z_d\}|z_s)\\
&&\times\left[\int_{0}^{\infty} d\mu_{max}\frac{dP}{d\mu_{max}}(\kappa,\gamma,z_d,z_s) \frac{f(\mu_{max}|\Delta m)N(>\frac{L_{lim}}{\mu_{var}},z_s)}{N(>\frac{L_{lim}}{\mu_{max}},z_s)}\right],
\end{eqnarray}
where $f(\mu_{max}|\Delta m)$ is the fraction of light-curves with $\mu_{max}$ that show variability above the level $\Delta m$. The integrals over $\mu_{max}$ in Eqns.~\ref{B} and \ref{F} were computed via Monte-Carlo. Integrating Eqn.~\ref{MLfrac} over source redshift we find
\begin{equation}
P(\Delta m,m_{lim})=\frac{\int_0^{\infty} dz_s\frac{dV_{cm}}{dz_s}N(>L_{lim},z_s)P(z_s,\Delta m,m_{lim})}{\int_0^{\infty} dz_s\frac{dV_{cm}}{dz_s}N(>L_{lim},z_s)},
\end{equation}
where $V_{cm}$ is comoving volume. For this calculation $P(z_s)$ was computed for $z_s=1$, 2, 3 and 4 (by $z_s=4$ Ly$\alpha$ has moved through B-band), and interpolated in $z_s$ during convolution with $N(>L_{lim},z_s)$. The resulting values of $P_{sgle}(\Delta m,m_{lim})$ (left), $P_{mult}(\Delta m,m_{lim})$ (center) and $P_{all}(\Delta m,m_{lim}) = P_{sgle}(\Delta m,m_{lim})+P_{mult}(\Delta m,m_{lim})$ (right) are plotted as a function of $\Delta m$ for several values of $m_{lim}$ in Fig.~\ref{fig7}. Also shown in Fig.~\ref{fig7} (grey lines) are the corresponding results for halos that are also composed of dark compact objects. 

Magnification bias introduces two competing effects with regard to the microlensing rate. First, low magnification images near the center of the lens, and those outside the Einstein radius are less likely to be observed, while the probability of observing images near the Einstein radius having values of $\kappa$ favorable to microlensing variability is enhanced. This leads to an increase in the microlensing rate. However magnification bias also reduces the fraction of images that are observed to undergo large fluctuations. The bias described in Eqn.~\ref{B} suppresses the frequency of observed large amplitude fluctuations because of the larger difference between the number of sources detectable during the survey at light-curve maximum and the number detectable while $\Delta m$ below the light-curve maximum. The net result is to increase the rate of stellar microlensing variability at a level $\Delta m>0.5$ by up to a factor of 10 (for $m_{lim}=17$) in images of both macro-lensed and singly imaged quasars. On the other hand, when we consider magnification bias in simulations where dark matter is composed of compact objects, we find the microlensing rates are not increased relative to the no bias case. In the case of multiple images, moderate optical depth is present right out to the critical radius for multiple imaging, so that there is no gain in terms of a more favorable $\kappa$. Thus the microlensing rate decreases with increased magnification bias, which is the opposite behavior to that of stars.

\begin{figure*}[hptb]
\epsscale{.4}
\plotone{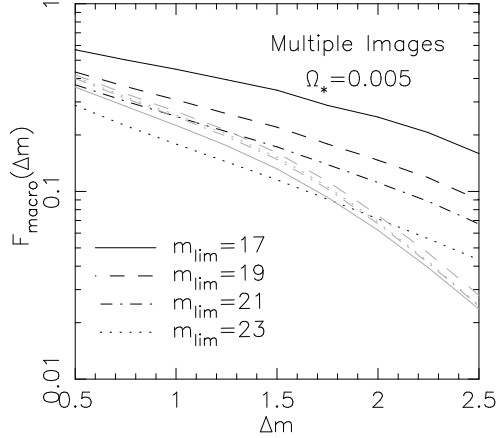}
\caption{Plots of the probability that a macrolensed image will be microlensed ($F_{mult}$) verses $\Delta M$ for monitoring of known lenses. The dark lines correspond to microlensing by stars, while the light lines assume dark matter to be in the form of compact objects. Values are shown for limiting B-magnitudes of $m_{lim}=17$, 19, 21 and 23.}
\label{fig8}
\end{figure*}

For our second example, we compute the rate of microlensing in the images of lensed quasars that have been previously identified in a survey for multiply imaged gravitationally lensed quasars. In this case the bias corresponding to the microlensing rate for images formed at a given set of microlensing parameters ($\kappa,\gamma$) is simply the bias for observing an image with that set of parameters in the initial survey. In general, this bias is computed using the sum of the magnifications of multiply imaged sources ($\mu_{tot})$ since the resolution of survey quality data is lower than that required to identify individual lensed images (e.g. Webster, Hewitt \& Irwin~1988). We compute the fraction of quasar images in macrolensed systems that exhibit microlensing. This is the most readily observed statistic since (following determination of a time delay) microlensing variability can be distinguished from intrinsic variability which is seen in all images. The fraction ($F_{mult}$) of macro-images (of quasars at redshift $z_s$) to undergo microlensing variability is
\begin{eqnarray}
\label{fmult}
\nonumber
&&\hspace{-10mm}F_{mult}(z_s,\Delta m,m_{lim}) =\\
&&\hspace{-10mm} \frac{\int_{10^{-4}}^{\infty} d\kappa\int_{0}^{z_s} dz_d p(\{\kappa,\gamma,\kappa_c,z_d\}|z_s)\left[\int_{0}^{\infty} d\mu_{tot}\frac{dP}{d\mu_{tot}}(\kappa,\gamma,z_d,z_s)f(\Delta m|\kappa,\gamma,z_d,z_s)N(>\frac{L_{lim}}{\mu_{tot}},z_s)\right]}{\int_{10^{-4}}^{\infty} d\kappa\int_{0}^{z_s} dz_d p(\{\kappa,\gamma,\kappa_c,z_d\}|z_s)\left[\int_{0}^{\infty} d\mu_{tot}\frac{dP}{d\mu_{tot}}(\kappa,\gamma,z_d,z_s)N(>\frac{L_{lim}}{\mu_{tot}},z_s)\right]}
\end{eqnarray}  
where $f(\Delta m|\kappa,\gamma,z_d,z_s)$ is the fraction of light-curves that vary by more than $\Delta m$ magnitudes, and $\frac{dP}{d\mu_{tot}}(\kappa,\gamma,z_d,z_s)$ is the normalized probability distribution for the sum of image magnifications (Sec.~\ref{magdist}). Integrating Eqn.~\ref{fmult} over source redshift we find
\begin{equation}
F_{mult}(\Delta m,m_{lim})=\frac{\int_{0}^{\infty} dz_s\frac{dV_{cm}}{dz_s}N(>L_{lim},z_s)F_{mult}(z_s,\Delta m,m_{lim})}{\int_{0}^{\infty} dz_s\frac{dV_{cm}}{dz_s}N(>L_{lim},z_s)}.
\end{equation}
Values of $F_{mult}$ are plotted as a function of $\Delta m$ for several values of $m_{lim}$ in Fig.~\ref{fig8}. Also shown in Fig.~\ref{fig8} (grey lines) are the corresponding results for halos that are composed of dark compact objects.

\begin{figure*}[hptb]
\epsscale{.7}
\plotone{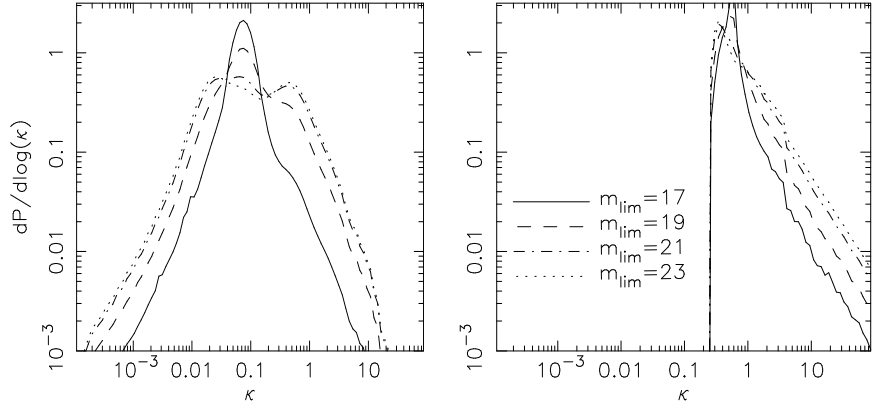}
\caption{Differential probability of $\kappa$ for multiply imaged sources in the presence of magnification bias. The left hand panel corresponds to microlensing by stars, while the right hand panel assumes dark matter to be in the form of compact objects. Distributions are shown for limiting B-magnitudes of $m_{lim}=17$, 19, 21 and 23. }
\label{fig9}
\end{figure*}

We find that $30-50\%$ of multiply imaged quasars should vary by more than $\Delta m=0.5$ magnitudes during a 10 year period. This rate drops by a factor of 5 for large amplitude fluctuations which is shallower than the previous example since the bias is not calculated from the light-curve. Note that these rates imply that microlensing induced variability is not uncommon in macrolensed quasars, even though the Einstein radius crossing time is many decades for most redshifts. This is because in the high $\kappa$ and $\gamma$ environments where most macro-images are found the caustic network has a typical scale-length significantly less than $\xi_o$. Furthermore, only part of a source need cross a caustic for significant variability to be observed. While the rate for microlensing by stars is very sensitive to magnification bias, the rates in the presence of dark compact objects in the dark matter halo are quite insensitive to the limiting magnitude. A large magnification bias results in an increased fraction of images near the galaxies Einstein radius where $\kappa_\star$ is typically around a 10th. Thus the microlensing rate is increased. On the other hand if the dark matter is composed of compact objects, the microlens optical depth is always greater than $0.25$ for multiple images. Hence magnification bias does not make much difference in this case. The rate does not drop appreciably with magnification bias, since unlike the previous example the bias was calculated independently of the light curve. The situation is described by Fig.~\ref{fig9} which shows the observed probability distribution for $\kappa$ assuming different limiting magnitudes. The left hand panel of Fig.~\ref{fig9} shows distributions assuming microlensing by stars, while the right hand panel assumes the halo is also composed of compact objects. Magnification bias imposes a sharp peak on the distribution near $\kappa=0.1$ for stars, increasing the microlensing rate since multiple images are less likely to be observed at low $\kappa$.

\subsection{The effect of microlensing on magnification bias and the distribution of flux ratios}
\label{bias}

We have seen that microlensing by stars affects the magnification distribution for a quasar image, as well as the distribution for the sum of the magnifications of multiply imaged quasars. Since microlensing in the two images is independent, we also expect microlensing to affect the flux ratios of images in macro-lensed systems. Furthermore, if microlensing affects the magnification distribution, then it might also qualitatively affect calculations of magnification bias for the fraction of multiply imaged quasars. As mentioned in the Introduction, Bartelmann \& Schneider~(1990) computed the effect of microlensing on the magnification bias and macro-image flux ratios. Under the assumptions of a point source, analytic forms for the magnification distributions (computed for different values of $\kappa$ and $\gamma$), and SIS galaxies having a fraction of density in compact objects with an outer radius, they concluded that microlensing could severely affect both the magnification bias and the flux ratio distribution. We are now in a position do better on all these scores, having computed magnification distributions numerically, for finite source size, and for a population of microlenses that have distributions resembling known stellar populations. 

\begin{figure*}[hptb]
\epsscale{.4}
\plotone{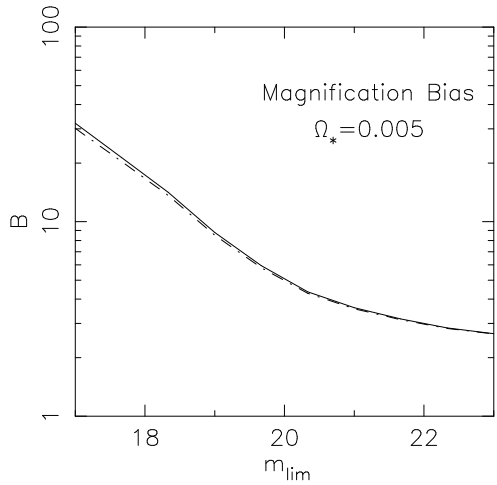}
\caption{Plots of the magnification bias $B$ verses limiting B-magnitude $m_{lim}$ The solid and dot-dashed lines correspond to biases including and not including microlensing.}
\label{fig10}
\end{figure*}

Using the luminosity function of Pei~(1995) we compute the magnification bias at source redshift $z_s$
\begin{equation}
\label{biaseqn}
B(z_s)=\frac{\int_0^{\infty} d\mu_{tot}\frac{dP}{d\mu_{tot}}(z_s)N(>\frac{L_{lim}}{\mu_{tot}},z_s)}{\int_0^{\infty} d\mu'[\tau_{mult}(z_s)\frac{dP(z_s)}{d\mu_{tot}}|_{\mu_{tot}=\mu'}+\frac{d\tau_{sing}}{d\mu'}]N(>\frac{L_{lim}}{\mu_{tot}},z_s)},
\end{equation}
which results in an overall bias of 
\begin{equation}
B=\frac{\int_0^{\infty} dz_s\tau(z_s)\frac{dV_{cm}}{dz_s}N(>L_{lim},z_s)B(z_s)}{\int_0^{\infty} dz_s\tau(z_s)\frac{dV_{cm}}{dz_s}N(>L_{lim},z_s)}.
\end{equation}
Note that Eqn.~\ref{biaseqn} includes the possibility that singly imaged quasars are magnified by microlenses beyond the galaxy Einstein radius. Fig.~\ref{fig10} shows the magnification bias for multiple imaging as a function of the limiting magnitude. The bias for a smooth SIS is shown for comparison (dot-dashed line). There is a slight increase in the bias if microlensing is assumed, however the effect is very small. Apparently additional bias due to the small increased probability for high magnifications is balanced by the possibility of having $\mu_{tot}<2$, as well as by the increased number of singly imaged sources (Bartelmann \& Schneider~1990). At redshifts where macrolensing is most likely, the distance ratio is of order 1, and so the source size is comparable to the projected size of the microlens Einstein radius. In this case the magnification distribution is quite narrow (Wambsganss~1992). Hence the excess of high magnifications formed by microlensed images is not as large as those predicted by the point source distribution, and as a result the effect on the macrolensing magnification bias is small, in contrast to the findings of Bartelmann \& Schneider~(1990). A similar result is found assuming dark matter to be composed of compact objects. 

\begin{figure*}[hptb]
\epsscale{.4}
\plotone{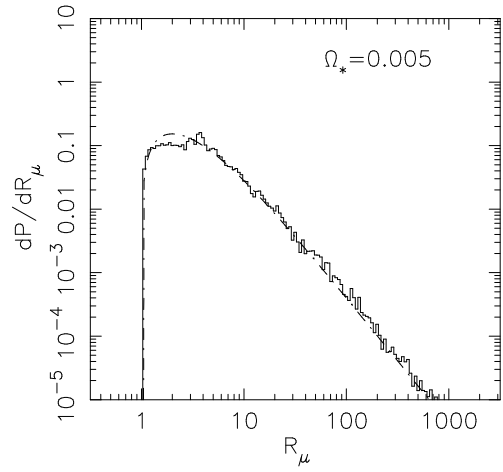}
\caption{Flux ratio ($\mu_1/\mu_2$) distributions for the images in multiply imaged quasars. The dot-dashed lines show the distribution for a smooth isothermal sphere and the source redshift was $z_s=3$.}
\label{fig11}
\end{figure*}

Finally, we compute the distribution of flux ratios $R\equiv\frac{\mu_1}{\mu_2}$ (where $\mu_1$ is now defined to be the image with the larger impact parameter, i.e. $1<x_1<2$ or $1/2>\kappa_1>1/4$). The distribution is shown in Fig.~\ref{fig11} for a source at $z_s=3$. The analytic distribution for a smooth SIS is shown for comparison. As a result of microlensing the likelihood of the distribution mode is lowered by a factor of $\sim2$, and a small excess of large flux-ratios is formed. This quantifies the often made statement that microlensing will influence flux ratios, and illustrates why optical flux ratios should not generally be used as constraints for models of gravitationally lensed galaxies.

\section{The Probability of Microlensing for Gamma Ray Burst Afterglows}
\label{GRB}

Just as microlensing promises to probe the central engines of quasars, microlensing of GRB afterglows offers a means to probe their structure (Loeb \& Perna~1998). Recently an anomalous event of 0.95 magnitudes was observed in the afterglow light-curve of GRB000301C and interpreted by Garnavich, Loeb and Stanek~(2000) as showing features consistent with the microlensing hypothesis (see also Gaudi, Granot \& Loeb~2001). With the afterglow light-curve of GRB000301C as motivation Koopmans \& Wambsganss~(2000) determined the a-posteriori probability that the afterglow was microlensed and found the probability to be small but not prohibitive. Furthermore, Mao \& Loeb~(2001) demonstrate that microlensing at the few percent level should be observed in all afterglows on timescales of a year (although they neglect clustering of microlenses). In this section we extend the calculation of Koopmans and Wambsganss~(2000) and consider the stellar population as microlenses (both alone and in addition to dark compact objects in the halo), as well as the fraction of microlensed GRB afterglows that will also be multiply imaged.

Our calculation follows the method described in Sec.~\ref{cross-sections}. However the GRB afterglow source is assumed stationary, while the relative motion is provided by its expansion. We assume the afterglow to be described by an expanding ring of constant surface brightness. The ring has a radius $R(t)=R_0(t/days)^{5/8}$ that expands as a function of time, and a width of $W\times R$ where $R_0$ is the width on day 1 (Waxman~1997). We take $R_0=3\times10^{16}(1+z_s)^{-\frac{5}{8}}$cm and $W=0.16$ (Loeb \& Perna~1998). The ring radius $R(t)$ must be adjusted for the smooth matter correction (Eqn.~\ref{defns}) as before. We assume that the afterglow is monitored for $30$ days (in the observers frame) following the burst. The luminosity function of GRB's from which the afterglows are identified is poorly known, and we do not consider magnification bias in the calculations of microlensing variability cross-sections. 
 
\begin{figure*}[hptb]
\epsscale{1.}
\plotone{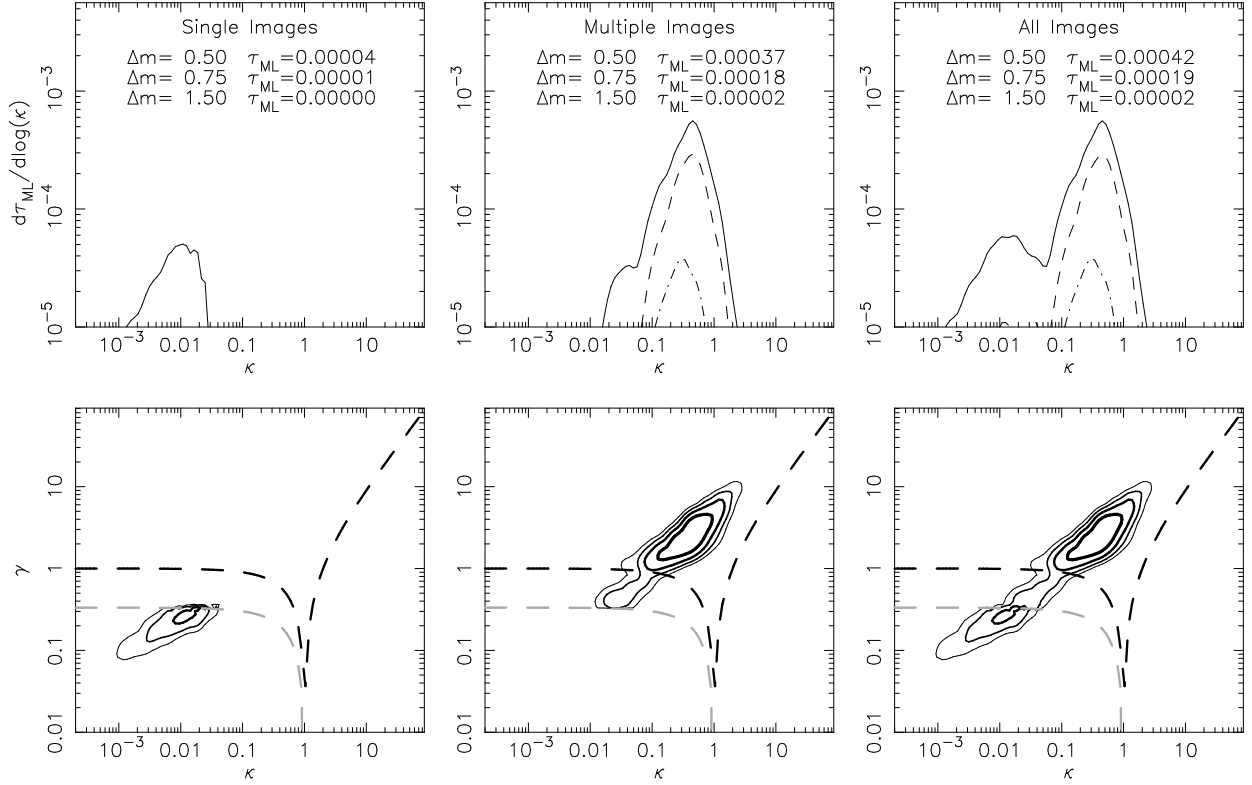}
\caption{Lower Rows: Contour plots (with contours spaced by multiples of $\sqrt{10}$) for the joint differential cross-section for $\kappa$ and $\gamma$ near the lines of sight to microlensed ($\Delta m>0.5$) GRB afterglows. The dashed lines show the condition for infinite magnification $1-\kappa=\pm\gamma$. Upper panels: Differential probability for $\kappa$ near the lines of sight to microlensed GRB afterglows. The solid, dashed and dot-dashed lines correspond to $\Delta m>0.5$, 0.75 and 1.5. The left, center and right panels show probabilities for single, multiple and all images respectively, the source redshift was $z_s=3$.}
\label{fig12}
\end{figure*}

The lower panels of Fig.~\ref{fig12} show $\frac{d^2\tau_{ML}}{d\kappa d\gamma}$, the conditional joint cross-section for values of $\kappa$ and $\gamma$ near lines of sight to microlensed images of GRB afterglows for the case of a source at $z_s=3$, and $\Delta m=0.5$ for singly imaged (left panel), multiply imaged (center panel, images are considered separately) and all (right panel) sources. Note that values of $\frac{d^2\tau_{ML}}{d\kappa d\gamma}$ for $\kappa<0.025$ were calculated by multiplying $\frac{d^2\tau_{ML}}{d\kappa d\gamma}|_{\kappa=0.025}$ by $(1-e^{-\kappa})^2/(1-e^{-0.025})^2$ (where the square is included because the motion is in both dimensions). The distribution shows many of the features seen in Figs.~\ref{fig1} \& \ref{fig4}, including depletion near the Einstein radius, demarcation between regions having singly and multiply (by the galaxy) imaged sources, and $\kappa<\gamma$ for all $\kappa$. 
The upper panels of Fig.~\ref{fig12} show $\frac{d\tau_{ML}}{d\kappa}$ obtained by integrating $\frac{d^2\tau_{ML}}{d\kappa d\gamma}$ over $\gamma$. Distributions are plotted for $\Delta m=0.5$, 1.0, 1.5, and 2.5 magnitudes. The microlensing cross-sections $\tau_{ML}(\Delta m,z_s)$ are also given. 

\begin{figure*}[hptb]
\epsscale{1.}
\plotone{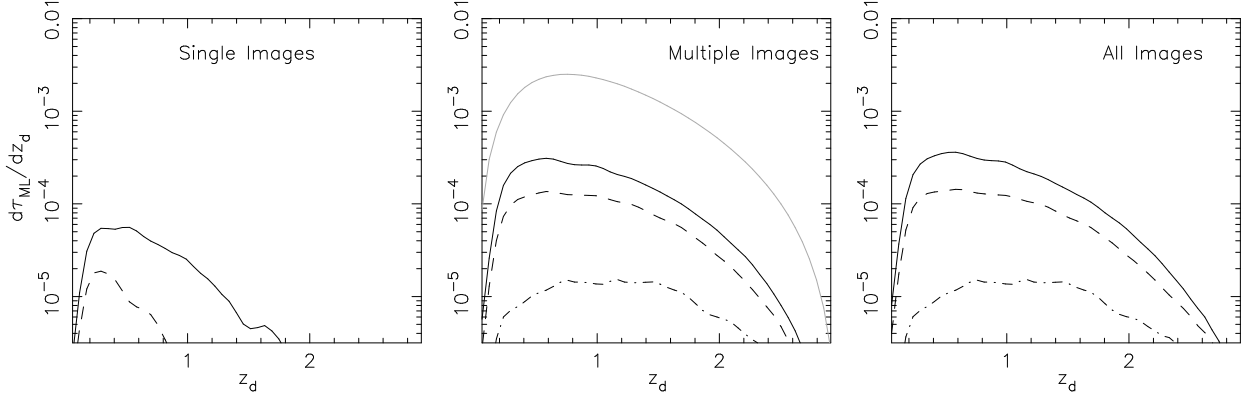}
\caption{Probability distributions for the redshift of galaxies responsible for microlensing of GRB afterglows. The left, center and right panels show values for single, multiple and all images respectively, and the solid, dashed and dot-dashed lines correspond to $\Delta m>0.5$, 0.75 and 1.5. The light line in the central panel shows the distribution of macrolens redshifts. The source redshift was $z_s=3$ .}
\label{fig13}
\end{figure*}

We found probability distributions for the redshift of galaxies whose stellar populations result in the microlensing of back-ground GRB afterglows. The resulting distributions $\frac{d\tau_{ML}}{dz_d}$ for the case of $z_s=3$ are plotted in Fig.~\ref{fig13} for singly imaged (left panel), multiply imaged (center panel) and all (right panel) sources. As before distributions are plotted for $\Delta m=0.5$, 1.0, 1.5, and 2.5 magnitudes, and the distribution of macrolens redshifts is included for comparison (light line). The shape of these distributions are similar to those for quasars. However, the distribution for singly imaged GRB afterglows is more peaked than that for quasars, and has a faster decline with redshift. 

\begin{figure*}[hptb]
\epsscale{1.}
\plotone{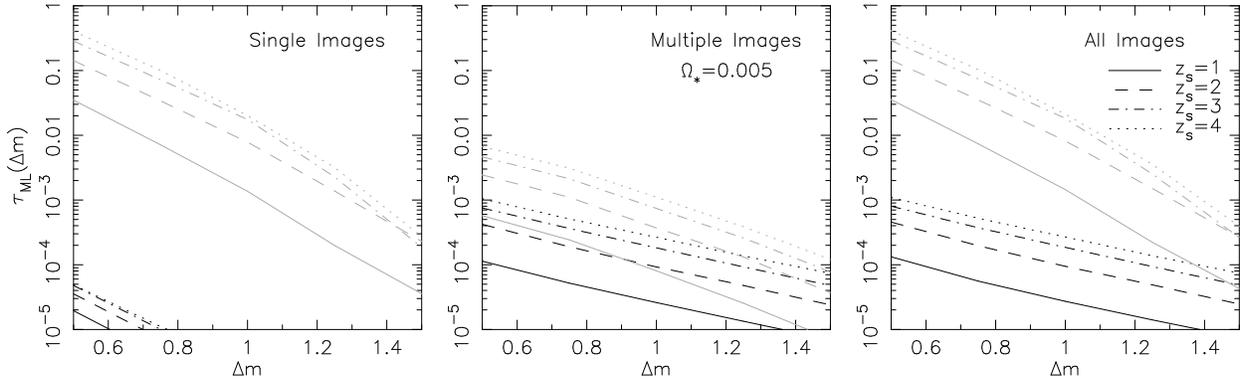}
\caption{Plots of the microlensing cross-section $\tau_{ML}$ verses $\Delta M$. The left, center and right panels show values for single, multiple and all images respectively. The dark lines correspond to microlensing by stars, while the light lines assume dark matter to be in the form of compact objects. Values are shown for $z_s=1$, 2, 3 and 4.}
\label{fig14}
\end{figure*}

We have computed the microlensing cross-section $\tau_{ML}(\Delta m,z_s)$ for GRB afterglows as a function of $\Delta m$ and plotted the results in Fig.~\ref{fig14} for $z_s=1$, 2, 3 and 4. The left, central and right panels show values for single, multiple (by the galaxy) and all images respectively. There are several points of interest. Firstly, the relative rates of microlensing between singly imaged and multiply imaged sources are similar for GRB afterglows and quasars. Around 1 in 1000 $z_s=3$ afterglows will be microlensed ($\Delta m=0.5$ magnitudes) by stars, however nearly all of these will be singly imaged. Secondly, the microlensing rate falls off steeply with the amplitude of the fluctuation $\Delta m$, dropping by a factor of $\sim20$ between $\Delta m=0.5$ and $\Delta m=1.5$. This is steeper than the corresponding dependency for quasars (see Fig.~\ref{fig5}). The largest fluctuations require a source that is small with respect to the projected Einstein radius, i.e. a low redshift for the galaxy containing the microlenses. For quasars this is accompanied by an increase in the effective transverse velocity, which helps to balance out the reduced probability for intercepting a caustic. However, the expansion velocity of the GRB afterglow is independent of $z_d$, hence the more rapid decline of microlensing rate with $\Delta m$. Fig.~\ref{fig14} also shows results obtained under the assumption that dark matter is composed of compact objects (light lines). We see an increase of a factor of $\sim10$ for macro-lensed afterglows, a larger discrepancy than for quasars. However, the fraction for all images ($\Delta M>0.5$ magnitudes) rises to more than 1 in 10. Note that we get a rate of $\sim1-3\%$ $(z_s>2)$ for $\Delta m>0.95$ magnitudes which is consistent with the value of 5\% found by Koopmans \& Wamsganss~(2001) for an optical depth of $\sim$0.25 over the whole sky. Features of the sort seen in the afterglow of GRB000301C should therefore be common if dark matter is composed of compact objects. This suggests that the large numbers of afterglows that are expected to be discovered by the upcoming \emph{swift} satellite will make an ideal probe of the cosmological density in dark compact objects since even a moderate number of featureless light-curves will provide a tight constraint. Furthermore, the high probability for GRB microlensing allows for the possibility of probing the distribution of compact objects in the halos around galaxies.

\begin{figure*}[hptb]
\epsscale{.4}
\plotone{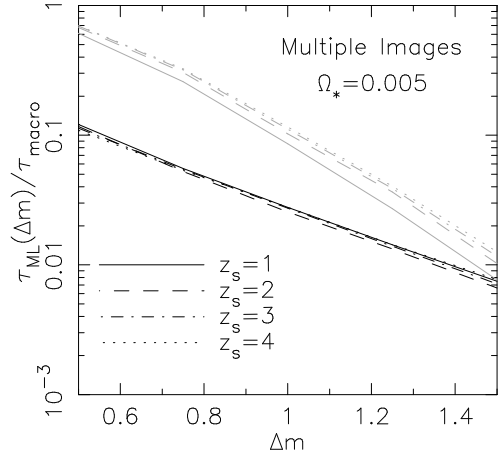}
\caption{Plots of the microlensing cross-section divided by the multiple imaging cross-section ($\tau_{ML}/\tau_{mult}$) verses $\Delta M$. The dark lines correspond to microlensing by stars, while the light lines assume dark matter to be in the form of compact objects. Values are shown for $z_s=1$, 2, 3 and 4.}
\label{fig15}
\end{figure*}

Finally we determine the fraction of multiply imaged GRB afterglows that are subject to microlensing. The results are shown in Fig.~\ref{fig15}. Only around 1 in 10 macrolensed GRB afterglow images will vary by more than 0.5 magnitudes due to stellar microlensing. This rate rises to around 60\% if the dark halos of galaxies are also composed of compact objects.

\section{Conclusions}

We have computed cosmological microlensing statistics for the populations of stars in elliptical/S0 galaxies. The stars were distributed with de-Vaucoleurs' profiles and embedded in a smooth dark halo to form an overall singular isothermal density distribution (SIS). Our calculation combines the joint cross-section for the microlensing optical depth $\kappa$ and shear $\gamma$ with a large number of numerical microlensing magnification patterns, which we assume resulted from microlenses each of $0.1M_{\odot}$. We calculated the fraction of quasar (with assumed source size of $10^{15}$cm) images that vary by more than $\Delta m$ magnitudes during 10 years of monitoring. We find that most microlensing will be observed in multiple images, and that the few cases of microlensing variability in sources that are not multiply imaged will primarily have $\gamma>0.1$ where the point-mass lens does not provide an adequate description. Furthermore, the majority of microlensed multiple images of quasars will be due to a lensing galaxy at a redshift lower than expected for macrolensing. In contrast, the rate of microlensing for sources that are not multiply imaged by the intervening galaxy is reasonably insensitive to the lens-galaxy redshift. 

We computed microlensing variability rates for two different examples of hypothetical surveys. first, for quasars in a continuous survey of a region of sky to a fixed limiting magnitude. For a limiting B-magnitude of $m_B=21$ we find that 1 quasar in 500 should vary by more than 0.5 magnitudes during 10 years of monitoring due to microlensing by stars. However 90\% of these are in multiply imaged systems. Macrolensed quasars therefore dominate microlensing statistics. If the dark halo (truncated so that the total mass density equals the critical density) is also composed of compact objects, then the fraction of quasar images subject to microlensing variability larger than 0.5 magnitudes rises to $\sim$10\%. On the other hand, the number of multiply imaged microlensed sources is quite insensitive to the inclusion of the additional compact objects. Therefore, if dark matter is composed of compact objects (as opposed to stars supplying all the microlenses) then microlensing is $\sim100$ times as common and is dominated by singly imaged quasars. The comparison of variability rates of lensed and unlensed quasars will therefore provide a powerful probe of the existence of dark compact objects. For our second example, we assumed that a sample of multiply imaged quasars had been previously selected as having the sum of the macro-images brighter than $m_B=21$. In this case we find that 1 image in 3 multiply imaged quasars should vary by more than 0.5 magnitudes during 10 years of monitoring. 

We have computed magnification distributions for individual quasar images in both single and multiple image systems, the magnification distribution for the sum of macrolensed images and the distribution of flux ratios for multiple image systems. Microlensing results in a spreading of all these distributions. In particular, in the presence of microlensing single images can have magnifications greater that 2, and the bright image of a multiply imaged quasar can have a magnification smaller than 2 (neither is formed by the SIS). From the distribution for the sum of the image magnifications we computed the magnification bias for the discovery of multiply imaged quasars that results from the inclusion of microlensing in the magnification distribution. Surprisingly, the effect on magnification bias for lens surveys is very small. However, we find that the inclusion of microlensing has a significant impact on the distribution of flux ratios, resulting in an excess of large values, and lowering the likeli-hood of the mode by a factor of 2 with respect to the smooth SIS distribution. This illustrates the point that optical flux-ratios make poor constraints for lens models.

We have also computed the fraction of GRB afterglow light-curves that exhibit microlensing. Specifically, we calculated the fraction of GRB afterglows that are perturbed by more than $\Delta m$ during the first 30 days. Qualitatively the results differ from those for quasars in that large amplitude fluctuations are much less common with respect to small amplitude fluctuations. We find that only 1 GRB afterglow in 1000 will vary by more than 0.5 magnitudes due to microlensing by stars. However most of these will also be multiply imaged by the galaxy. If the halo is comprised of dark compact objects we find that the fraction of GRB afterglows that vary by more than $\Delta M>0.5$ magnitudes rises to more than 1 in 10. Features of the sort seen in the afterglow of GRB000301C (but with a smaller amplitude) should therefore be very common if dark matter is composed of compact objects. If a GRB afterglow is multiply imaged, then we find that there is $\sim$1 chance in 10 that microlensing by stars will also be observed at a level greater than 0.5 magnitudes, but a 60\% chance if the dark matter is also in compact objects.

\acknowledgements
The authors would like to thank Joachim Wambsganss for providing his \emph{microlens} ray-tracing program with which the microlensing simulations presented in this paper were computed. This work was supported by NASA through a Hubble Fellowship grant from the Space Telescope Science Institute, which is operated by the Association of Universities for Research in Astronomy, Inc., under NASA contract NAS 5-26555 (for J.S.B.W.), and by NSG grant NAG5-9274 to ELT.

\newpage

\end{document}